\documentclass[]{iopart}
\usepackage{cite} 
\usepackage{iopams}
\usepackage{epsfig}
\usepackage{bm}
\usepackage{longtable}
\usepackage{graphicx}
\usepackage{graphics}
\usepackage{dcolumn}
\usepackage{color}

\begin{document}

\title[Contact interactions for CI]{On the renormalization 
of contact interactions 
for the configuration-interaction method in two dimensions}

\author{Massimo Rontani$^1$, G Eriksson$^2$, S {{\AA}berg}$^2$
and S M Reimann$^2$}
\eads{\mailto{massimo.rontani@nano.cnr.it}}
\address{$^1$ CNR-NANO, Via Campi 213a, 41125 Modena, Italy}
\address{$^2$ Mathematical Physics, LTH, Lund University,    
P. O. Box 118, 22100 Lund, Sweden}





\begin{abstract}
The configuration interaction (CI) method
for calculating the exact eigenstates of a quantum-mechanical few-body system 
is problematic when applied to particles interacting through contact forces.
In dimensions higher than one the approach fails due to the  
pathology of the Dirac $\delta$-potential, making it impossible to reach 
convergence by gradually increasing the size of the Hilbert space. 
However, this problem may be cured in a rather simple manner 
by renormalizing the strength of the contact potential when 
diagonalizing in a truncated Hilbert space.  One hereby relies on the 
comparison of the CI results to the two-body ground-state
energy obtained by the exact solution of the 
Schr\"odinger equation for a regularized contact interaction.
We here discuss a scheme that provides cutoff-independent few-body physical 
observables. The method is applied to a few-body system of ultracold atoms 
confined by a two-dimensional harmonic oscillator.
\end{abstract}

\pacs{67.85.-d, 31.15.ac, 67.85.Lm, 03.75.Ss}
\submitto{\JPB}

\maketitle

\section{Introduction}

Trapped ultracold atom 
gases \cite{Dalfovo1999,Leggett2001,PethickSmith2002,PitaevskiiStringari2003,Leggett2006,Bloch2008,Giorgini2008} are an ideal laboratory for the investigation
of many-body physical phenomena.
From a theoretical perspective, a major issue is the complexity 
of the atom-atom interaction potential which in numerical 
simulations can hardly be considered in its full form.  
However, one often is only interested in the low-energy scattering 
properties which in many cases may be accounted for by simple pseudopotentials. 
A standard approximation to model the atom-atom interactions in an ultracold 
gas of fermionic  or bosonic atoms is a contact potential \cite{Huang} 
describing the $s$-wave isotropic scattering  
by means of only one parameter, the scattering length $a$. 
For cold atom gases
this pseudopotential has been widely and successfully used 
(see for example references \cite{Dalfovo1999,Leggett2001,PethickSmith2002,PitaevskiiStringari2003,Leggett2006,Bloch2008,Giorgini2008}).
However, the use of such pointlike interactions is mathematically 
troublesome \cite{Huang}. 
In their simple form of Dirac $\delta$-functions contact
potentials are unphysical in two and three
dimensions and one needs to introduce
a proper regularization \cite{Galitskii1958,Gorkov1961,Leggett1980bis,adhikari1997,ghosh1998,Camblong2002,Pricoupenko06,suto,Busch,Cabo,Esry,Castin,Pricoupenko2011}.
In addition, there are subtle drawbacks which become
particularly relevant when trying to solve the many-particle problem exactly.
The crux of the matter (as noted by Huang \cite{Huang}) lies
in the fact that the regularized operators  may be  
non-Hermitian, which prohibits finding their eigenvalues
by variational methods. In particular, the direct diagonalization
of the Hamiltonian on a {\em complete} space 
cannot be applied \cite{Huang,Busch,Doganov2013},
unless the class of allowed basis wave functions 
obeys special (and often impractical) boundary conditions \cite{Esry}.
This also invalidates the direct application of the contact potential 
in the so-called configuration interaction (CI) approach 
\footnote{Other
model interactions have been considered in the CI literature, such as 
Gaussian 
\cite{Blume2007,Stecher2007bis,Stecher2007ter,vonStecher2008,Christensson2009} 
and Morse \cite{Esry} potentials.}
(also known as the method of `exact diagonalization'). 
In the limit of small particle numbers and not too strong interactions 
the CI approach would otherwise allow an accurate treatment of correlation 
effects, also giving access to excited 
states \cite{McWeeny,Helgaker,Jensen,Brasken2000,ReimannRMP,RontaniCI}.
It is therefore desirable to develop a  
scheme that can incorporate the regularization in a simple 
and straightforward manner.

In this paper we discuss a renormalization procedure for few-body 
systems with contact interactions in two dimensions (2D). 
For the confinement we choose the example of a 2D harmonic oscillator, 
a system that has been widely investigated in the context of ultracold 
atomic gases \cite{Fetter2009} and whose exact two-body states 
are known \cite{Busch}.
From a computational point of view, the 2D case is simpler than the
three-dimensional one, 
however providing a similar complexity of the problem at hand. 
The renormalization procedure that we discuss here is similar 
in spirit to the ultraviolet 
renormalization adopted in analogous contexts 
by means of diverse techniques, 
including space discretization \cite{Castinbis,Drut2013}, 
dimensional \cite{Camblong2002} and momentum-cutoff 
regularization \cite{Esry,Camblong2002,Braaten1997,Bulgac2006,Stecher2007,Stetcu2007a,Stetcu2007b,Alhassid2008,Zinner2009,Gilbreth2012,Tolle2013,Gilbreth2013}, 
and other 
approaches \cite{Bulgac2002,Pricoupenko2004,Johnson2010,Boettcher2012,Yan2015}.
We demonstrate explicitly how in a truncated Hilbert subspace 
with properly renormalized interactions the CI algorithm correctly 
reproduces the exact energies and wave functions of both ground and 
excited states, except when the relative distance between 
the particles becomes too small. The coupling constant of the 
contact potential employed in the calculation can be mapped onto 
the two-dimensional scattering length $a$. However, this mapping 
significantly depends on the energy cutoff. A relation between 
these quantities is established for the case of two particles 
confined in a 2D harmonic trap, for which the exact regularized 
solutions are available \cite{Busch}. For three particles, 
the method is further validated through a comparison of semi-analytical 
results \cite{Drummond} with the CI data.
The procedure can then also be applied in the case of larger systems, 
where such analytic solutions do not exist. As a case study for the latter,  
we further compute the low-lying energy spectrum of $N$ fermions confined
in a harmonic trap, with $N \le 5$. 
The CI results suggest that when diagonalizing in a truncated Hilbert 
space, by adapting the coupling constant one can accurately evaluate 
cutoff-independent $N$-body physical observables.  

\section{The two-body problem treated in relative coordinates}
\label{s:two}

In order to investigate the nature of the 
contact interactions let us begin with the simple case of two interacting particles in a 
two-dimensional harmonic trap of  frequency $\omega _0$.
Throughout this paper we use as energy unit $\hbar \omega _0$ and as length
unit the oscillator length, $\ell _0= (\hbar /m\omega_0)^{1/2}$. 
The two-body Hamiltonian, conveniently 
written in the center-of-mass and relative-motion
coordinates $\bm{R}=(\bm{r}_1+\bm{r}_2)/2$ and $\bm{r}=\bm{r}_1
-\bm{r}_2$, respectively, then reads as
\begin{equation}
H_2 = H_{{\rm com}}+H_{{\rm rel}},
\label{eq:H2}
\end{equation}
with
\begin{equation}
 H_{{\rm com}} = -\frac{1}{4}\nabla_{\bm{R}}^2+R^2
\end{equation}
and
\begin{equation}
H_{{\rm rel}} = -\nabla_{\bm{r}}^2
+\frac{1}{4}r^2+g\,\delta({\bm{r}}) 
\label{eq:Hrel}
\end{equation}
where $g$ is the coupling constant 
in units of $(\hbar \omega _0) \ell _0^2 $.
While the center-of-mass eigenstates of $H_{{\rm com}}$ are simply
(2D) harmonic oscillator orbitals in the $\bm{R}$ coordinate
with trivial integer energies $E_{{\rm com}}$, 
the non-trivial part of the two-body Hamiltonian, $H_{{\rm rel}}$, 
pertains to the relative motion. 
If the relative-motion angular momentum $M_{{\rm rel}}$
is even (odd), the solution equally
holds for spinless bosons and spinful fermions with total spin zero (one).
Here we investigate the low-lying singlet states of either 
two spinless bosons
or two fermions of opposite spin 1/2.

Solving the eigenvalue problem for $M_{{\rm rel}}=0$
one diagonalizes the matrix
\begin{equation}
\left<n^{\prime}\right|H_{{\rm rel}}
\left|n\right>=
(2n+1)\delta_{n,n^{\prime}}+\frac{g}{2\pi},
\label{eq:matrix}
\end{equation}
where $\left|n\right>= \varphi_n(\bm{r}) \equiv \varphi_{n0}(\bm{r})$ 
stands for the 2D oscillator orbital in the
$\bm{r}$-space with $n$ nodes in the radial direction
and angular momentum $m=0$ (the other sectors with
$M_{{\rm rel}}\neq 0$ are trivial since off-diagonal
matrix elements are zero). 
The explicit orbital form is
$\varphi_{n}(r) = (2\pi)^{-1/2} L_n(r^2/2)e^{-r^2/4}$, where $L_n(z)$ is
the Laguerre polynomial of order $n$.
In practice, one must restrict the number of basis functions, $N_b$, 
with $N_b = n_{{\rm max}}+1$ and $n_{{\rm max}}$ being the number of nodes
of the highest-energy orbital.
For any $g<0$, the ground-state energy obtained from the
direct diagonalization of the matrix (\ref{eq:matrix}) does not converge but 
decreases monotonously as the basis size $N_b$
increases. For $g>0$, however, the energy converges to the 
noninteracting value (i.e., unity) \cite{suto}.
Such behavior could also be expected from the fact that the 
offdiagonal Hamiltonian matrix elements are independent from $n$, 
as shown by (\ref{eq:matrix}) \footnote{Considering
the limiting case of very large interaction
strength $\left|g\right|$, the Hamiltonian can be approximately written in
matrix form as $H_{{\rm rel}} \approx \bm{I}+g/(2\pi) \bm{1}$, where
$\bm{I}$ is the identity and
$\bm{1}$ is the $N_b\times N_b$ matrix with all elements equal to one.
For large values of $N_b$, for $g\gg 0$ the lowest eigenvalue is unity, i.e.,
the contact interaction does not provide any scattering. For $g\ll 0$,
one obtains $-\left|g\right|N_b/(2\pi)$, i.e., the ground-state energy
diverges with $N_b$ for attractive interaction \cite{suto}.
Note that in the well behaved one-dimensional case the
off-diagonal element is either zero or it decreases with $n$ as
$\sim (-1)^{(n+n^{\prime})/2} (nn^{\prime})^{-1/4}$.
}.
This clearly shows the pathology of the contact interaction 
in dimensions higher than one.

\subsection{Exact solution}
\label{s:twoA}

A proper solution of the two-body problem
consists in introducing a regularized contact potential. This path was 
followed by Busch and coworkers \cite{Busch}, whose results we partly 
recall in this section.  Olshanii and Pricoupenko \cite{Olshanii} 
provided a general form of the pseudopotential, $V_{{\rm pseudo}}$,
which should replace the simple Dirac $\delta$-function in $H_{{\rm rel}}$,
\begin{equation}
V_{{\rm pseudo}} =
-\frac{2\pi\delta(\bm{r})}{\ln({\cal{A}}a\Lambda)}
\left[1-\ln({\cal{A}}\Lambda r)\,r
\frac{\partial}{\partial r}\right]_{r \rightarrow 0^+},
\label{eq:regularized}
\end{equation}
where $\Lambda$ is an arbitrary constant, $a$ 
is the two-dimensional
scattering length, ${\cal{A}}=e^{\gamma}/2$, and 
$\gamma = 0.5772\ldots$ is the Euler-Mascheroni constant.

The eigenstates of the regularized form of the relative-motion Hamiltonian,
$H_{{\rm rel}}^{{\rm reg}}$, with
\begin{equation}
H_{{\rm rel}}^{{\rm reg}} = -\nabla_{\bm{r}}^2
+\frac{1}{4}r^2+ V_{{\rm pseudo}},
\label{eq:Hreg}
\end{equation}
are obtained by imposing that the
wave functions, $\Psi(\bm{r})$, written
as linear superpositions of the orbitals $\varphi_n(\bm{r})$
with unknown coefficients $c_n$,
\begin{equation}
\Psi(\bm{r})=\sum_{n=0}^{\infty}c_n\,\varphi_n(\bm{r}),
\label{eq:infsum}
\end{equation}
must solve the eigenvalue problem
\begin{equation}
H_{{\rm rel}}^{{\rm reg}} \,\Psi(\bm{r}) = 
E_{{\rm rel}}^{{\rm reg}}  \,  \Psi(\bm{r}).
\label{eq:eigenproblem}
\end{equation}
Equations (\ref{eq:infsum}) and (\ref{eq:eigenproblem}), 
together with the requirement
of normalization, determine
both the coefficients $c_n$ and the energy $E_{{\rm rel}}^{{\rm reg}} $,
the latter through the equation
\begin{equation}
\psi(1/2 - E_{{\rm rel}}^{{\rm reg}} /2)=\ln(2/a^2),
\label{eq:nosbusch}
\end{equation}
where $\psi(z)$ is the digamma function of argument $z$ \footnote{Equation
 (\ref{eq:nosbusch}) actually differs from (21) of reference 
\cite{Busch} for a numerical factor which is due to the 
different definition of $a$. Our usage here is consistent with 
the pseudopotential definition (\ref{eq:regularized}).
}.

\begin{figure}[htbp]
\setlength{\unitlength}{1 cm}
\begin{indented}
\item[]\begin{picture}(8.5,6.5)
\put(0.2,-0.2){\epsfig{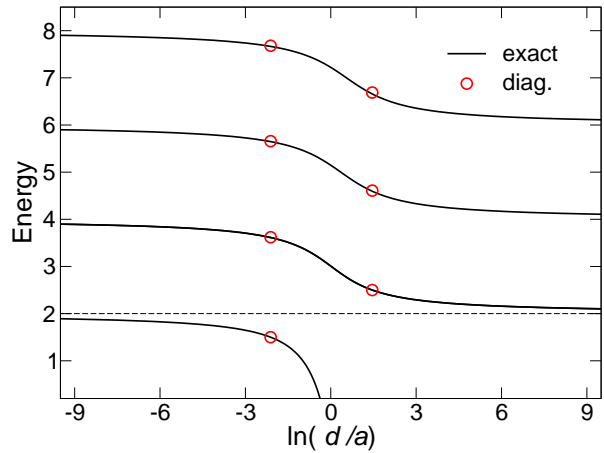}}
\end{picture}
\end{indented}
\caption{(color online).
Exact energy spectrum $E^{{\rm reg}}$
of two particles (either fermions with total spin zero
or spinless bosons) versus two-dimensional scattering length $a$.
The circles show the energies obtained by direct diagonalization 
given in table \ref{table1} and the dashed line is the 
noninteracting ground-state energy.
$d=\sqrt{2}$ is the harmonic-oscillator length
in the relative frame and $M_{{\rm rel}} = 0$.
}
\label{f:E2}
\end{figure}

The total energy $E^{{\rm reg}} = E_{{\rm rel}}^{{\rm reg}}
+ E_{{\rm com}}$ is plotted 
as a function of $\ln{(d/a)}$
in figure \ref{f:E2}, with $d=\sqrt{2}$ being the harmonic
oscillator length in the relative frame
(the scattering length $a$ is always positive in two dimensions,
as discussed e.g.~in Ref.~\cite{Petrov2001}).
The wave function is
\begin{equation}
\Psi(\bm{r})\propto U(-\nu,1,r^2/2)e^{-r^2/4},
\label{eq:Psiexact}
\end{equation}
where $\nu = (E_{{\rm rel}}^{{\rm reg}} -1)/2$ and $U(a,b,z)$ is the 
confluent hypergeometric function.

The key point in the derivation of
(\ref{eq:nosbusch}) is that one is not allowed to interchange
the limiting procedure contained in the definition (\ref{eq:regularized})
of $V_{{\rm pseudo}}$,
$\lim_{r\rightarrow 0^+}\partial/\partial r$, with the infinite
summation appearing in (\ref{eq:infsum}), $\sum_{n=0}^{\infty}$.
In fact, by doing so, the `regularizing' part of the
pseudopotential,
$\lim_{r\rightarrow 0^+} \ln({\cal{A}}\Lambda r)\,r\, \partial/\partial r$,
provides a null result when applied to each addendum $\varphi_n(\bm{r})$
of the sum, which is well behaved at the origin. In this case, the only
possible solution of (\ref{eq:eigenproblem}) is the noninteracting one.
For the same reason, the regularization of $V_{{\rm pseudo}}$
is useless if the expansion (\ref{eq:infsum})
is finite, as it is of course always the case when diagonalizing numerically, 
making the regularization procedure irrelevant in this case. 

\subsection{Diagonalization in a finite Hilbert space}
\label{s:twoB}

Let us now investigate whether it is possible to perform a numerical diagonalization with the bare $\delta $-interaction in a truncated Hilbert space and to relate the result to the regularized solution. 
We here still restrict ourselves to the two-body case.
The eigenvalues are obtained by performing the diagonalization
in a given basis set, i.e., for a given value of $N_b$,  
at a fixed value of the coupling constant $g$. 
The result naturally depends on the number $N_b$ of orbitals used.
In order to obtain results which nevertheless can be interpreted physically
we suggest to
\newline (i) Fix the Hilbert space size for the two-body system, i.e., $N_b$.
\newline (ii) Link the value of the coupling constant $g$ 
to be used in the direct diagonalization 
to the physically meaningful value of the exact, fully regularized two-body ground-state (GS) energy
for a given scattering length $a$, 
\begin{equation}
\mathcal{E} \equiv E_{\rm GS}^{{\rm reg}}(a) = 
E_{{\rm GS}}^{{\rm diag}}(g, N_b).
\label{eq:constraint}
\end{equation}
Here $E_{{\rm GS}}^{{\rm diag}}(g, N_b)$  
is the energy obtained by the direct diagonalization 
for a specific value of $g$ in a given basis 
set with fixed  $N_b$, and $E_{\rm GS}^{{\rm reg}}(a)$
is the desired regularized two-body ground-state energy that we 
in the following denote as $\mathcal{E}$
(the value $\mathcal{E}=2$ is the energy in the noninteracting case, $g=0$).

In the following, we evaluate the above procedure by comparing 
the results obtained through direct diagonalization with the exact 
and fully regularized results 
that are available in the literature, i.e.,
energies and wave functions of two-body excited states \cite{Busch} 
(Sec.~\ref{s:twoC})
as well as exact eigenstates for three particles \cite{Drummond} 
(Secs.~\ref{s:threeA}
and \ref{s:threeB}). 
This also allows to estimate the error associated with the truncation
of the Hilbert space size.

\subsection{Energies and wave functions of two-body excited states}
\label{s:twoC}

For two particles described in relative
coordinates, let us now fix the Hilbert space size through the 
number of relative orbitals $N_b$ and consider two cases, 
corresponding to repulsive and attractive 
interaction, respectively. 
As a specific example, we set 
$N_b=13$. Within this restricted space, we then chose as an example 
the two-body GS energies 
$\mathcal{E} = 1.500$ and $\mathcal{E} = 2.500$. 
In the direct diagonalization of 
(\ref{eq:matrix}) these energies are obtained with  
the coupling constants $g=-1.855$ and $g=30.78$, respectively.

Note from figure \ref{f:E2} that 
$\ln{(d/a)}$ assumes both positive
and negative values, with the noninteracting
ground-state energy (dashed line) being reached at 
both infinities on the real axis.
Therefore,
moving away from the $\mathcal{E} = 2$
asymptotic value at $\ln{(d/a)}=\pm \infty$ 
by increasing 
(decreasing) continuously the energy allows us to
identify the branch that physically corresponds to repulsive (attractive)
interaction, i.e., $g>0$ ($g<0$).  
Solving (\ref{eq:constraint}) numerically for $\mathcal{E} = 1.500$
and $\mathcal{E} = 2.500$ we respectively obtain $a = 11.71$ 
and  $a=0.3294$ (cf.~the two lowest-lying circles in figure \ref{f:E2}). 

Let us first compare the energies and
wave functions of the two-body excited states obtained 
by direct diagonalization 
of (\ref{eq:matrix}) to the
analogous exact and properly regularized quantities 
obtained from the knowledge of $a$ through (\ref{eq:nosbusch}) and
(\ref{eq:Psiexact}). 
\begin{table}
\caption{\label{table1} Comparison between exact regularized and 
directly diagonalized 
two-body energies for both ground and excited states, 
for attractive (2nd and 3rd column) and repulsive (4th and 5th
column) interaction, respectively. The exact regularized 
(`exact') and the diagonalized (`diag.') 
data are linked
by matching the corresponding GS energies, providing $(a,g)=
(11.71,-1.855)$ and $(0.3294,30.78)$ for $\mathcal{E}=1.500$ and  
$2.500$, respectively,
where $a$ is obtained by solving (\ref{eq:nosbusch})
and $g$ is obtained by diagonalizing (\ref{eq:matrix})
with $N_b=13$. 
}
\begin{indented}
\item[]\begin{tabular}{ccccc}
\br
& \multicolumn{2}{c}{ $g < 0$ } & \multicolumn{2}{c}{ $g > 0$ } \\

${\rm level}$ & {\rm exact} & {\rm diag.} & {\rm exact} & {\rm diag.} 
\\ 
\mr
\\
{\rm ground state} & 1.500 & 1.500 & 2.500 & 2.500  \\
{\rm 1st excited} & 3.613 & 3.619 & 4.596 & 4.608  \\
{\rm  2nd excited} & 5.647 & 5.657 & 6.658 & 6.688  \\
{\rm  3rd excited} & 7.666 & 7.680 & 8.706 & 8.757  \\
\br
\end{tabular}
\end{indented}
\end{table}

\begin{figure}
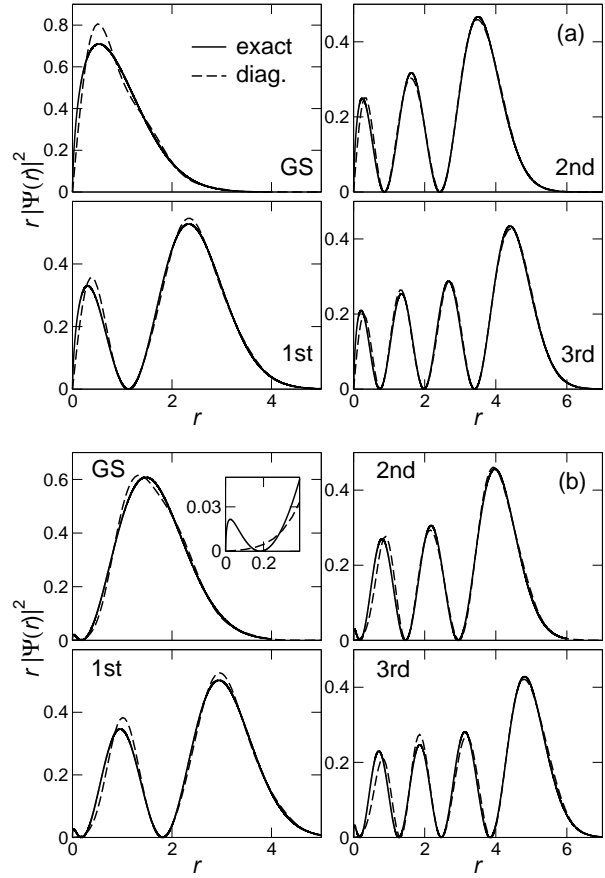

\begin{indented}
\item[]
\vspace{5mm}
\centerline{\epsfig{file=figure2a.eps,width=3.1in,,angle=0}}
\vspace{3mm}
\centerline{\epsfig{file=figure2b.eps,width=3.1in,,angle=0}}
\end{indented}
\caption{
Radial probability density $r\left|\Psi(r)\right|^2$ 
of finding two particles at distance $r$,
from both the regularized exact solution (solid lines) 
and the direct diagonalization  for fixed $N_b$ and 
renormalized coupling constant $g$ (dashed lines), 
for attractive (a) and repulsive (b) interaction, respectively.
The panels display the ground (GS), first- (1st), second- (2nd),
and third-excited (3rd) states.
Here $\Psi(r)$ is the relative-motion wave function
normalized as
$\int dr\,r
\left|\Psi(r)\right|^2=1$ for the exact regularized and 
the directly diagonalized solutions, respectively. 
The GS energy $\mathcal{E}$ is 1.500 and 2.500
implying scattering lengths
$a=11.71$ and 0.3294 for
attractive (a) and repulsive (b) interaction,
respectively, whereas the renormalized coupling constant used in the 
CI diagonalization is
$g=-1.855$ and 30.78, respectively, corresponding 
to a single-particle basis
of $N_b=13$ orbitals.
The energies of the states considered here are displayed in 
the 2nd and 3rd (4th and 5th) columns of table
\ref{table1} for attractive and repulsive interaction,
respectively, as well as in figure \ref{f:E2}.
The inset in the GS panel of figure (b) is a blow-up close to the origin.} 
\label{excited_r}
\end{figure}

This is done in figure \ref{f:E2} and table \ref{table1} 
that compare energy levels up to the third excited
state. For the chosen coupling strength $g$, 
the energies obtained by direct diagonalization 
(circles in figure \ref{f:E2}) nicely
match the regularized exact solution (solid lines),
the worst relative error being only 7 parts per thousand (for the 3rd
excited level in the repulsive case, cf.~table \ref{table1}) 
for our example of $N_b=13$. 
This very good agreement is
confirmed by the wave function analysis, 
reported in figures \ref{excited_r}(a)
and (b) for 
attractive and repulsive interaction, respectively.

The plots compare the 
radial probability densities of finding two
particles at distance $r$, up to the third excited state.
We see in both attractive and repulsive cases 
(figures \ref{excited_r}(a) and (b), respectively) that the overlap
of the probability densities as well as the agreement 
regarding node locations
is almost perfect far from the origin, but 
becomes progressively worse as $r\rightarrow 0$. 
In fact, the square modulus of the exact regularized wave function
presents a logarithmic singularity at the origin, originating from the
behavior of the hypergeometric function of (\ref{eq:Psiexact}):
\begin{displaymath}
\Psi (r) \approx \frac{A}{\Gamma(-\nu)}\left[\ln{r^2} 
+\psi(-\nu)\right]\quad{\rm for}\quad r\approx 0,
\end{displaymath}
where $A$ is a normalization constant.
Note the appearance of a node in the exact ground-state wave function 
for repulsive interaction [cf.~inset of
figure \ref{excited_r}(b)], due to the occurrence of a strongly bound 
molecular state
that is much deeper in energy \footnote{This low-lying 
energy branch of strongly bound molecular dimers is 
the absolute ground state.}.
Any expansion over a finite set of basis functions 
regular at the origin is unable
to reproduce the logarithmic singularity.  
However, in 2D this is expected to have little effect 
on the calculations of observables dominated by the whole range of $r$.

\subsection{Truncation of the Hilbert space and short-distance cutoff}
\label{s:twoD}

\begin{figure}[htbp]
\setlength{\unitlength}{1 cm}
\begin{indented}
\item[]\begin{picture}(8.5,5.5)
\put(0.2,-0.2){\epsfig{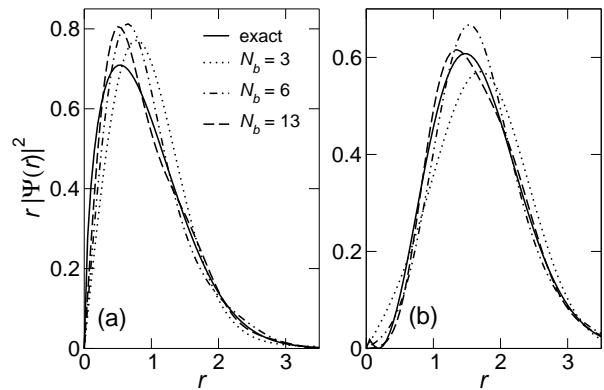}}
\end{picture}
\end{indented}
\caption{
Radial probability density $r\left|\Psi(r)\right|^2$
of finding two particles at distance $r$
obtained from the direct diagonalization of (\ref{eq:matrix})
over subspaces
of increasing $N_b$. Here $\Psi(r)$ is the relative-motion wave function
normalized as $\int dr\,r \left|\Psi(r)\right|^2=1$.
The solid lines are the data from the regularized 
exact solution corresponding to 
$a=11.71$ and 0.3294 for attractive (a) and repulsive (b) interaction,
respectively. The values of the coupling constants $g$ 
chosen in the direct diagonalization change according 
to the different single-particle
basis sizes $N_b$, providing as GS energies 
respectively $\mathcal{E}=1.500$ (a) and $\mathcal{E}=2.500$ (b). 
}
\label{f:cutoffs}
\end{figure}

To understand the consequences on the wave function
let us take a look at the evolution of the ground state 
obtained through direct diagonalization when the parameter
$N_b$ (controlling the energy cutoff in the single particle 
basis in relative coordinates \cite{Coon2012,Konig2014}) 
is increased, going from dotted to dashed curves in 
figure \ref{f:cutoffs}, respectively.  
We see the
same trend for both attractive (a) and
repulsive (b) interactions, with the weight of 
$r \left|\Psi_{{\rm diag}}(r)\right|^2$ from the direct-diagonalization result 
shifting towards the origin.
This increases the overlap with the regularized exact radial density
$r\left|\Psi_{{\rm exact}}(r)\right|^2$.
This behavior suggests that the effect of the
 cutoff in the direct diagonalization is to provide
a complementary short-distance cutoff $r_c$ in real space, the 
larger $N_b$ (higher energy cut-off) the smaller $r_c$ 
(better resolution).

To be quantitative,
we define the real-space cutoff $r_c$ as the lower bound of
the interval $(r_c,+\infty)$ within which the wave functions 
obtained by the renormalized numerical diagonalization 
and the exact analytic regularization largely overlap.
In detail, we define the integral function $D(r)$
as the deviation of the CI probability density $r
\left|\Psi_{{\rm diag}}(r)\right|^2$
from its exact counterpart $r\left|\Psi_{{\rm exact}}(r)\right|^2$,
which is provided by integration between $r$ and $\infty$:
\begin{equation}
D(r)=\int_r^{+\infty}dr'\left|\left|\Psi_{{\rm diag}}(r')\right|^2-
\left|\Psi_{{\rm exact}}(r')\right|^2 \right|r'.
\end{equation}
It is important to note that,
since both radial probabilities are separately
normalized to one, then $0\le D(r) \le 1$ and $D(r)\rightarrow 0$ for 
$r\rightarrow 0$.
Furthermore, $D(r)$ vanishes for $r\rightarrow \infty$ since
both wave functions decay at infinity. 
From these properties the generic behavior of $D(r)$
is as follows.
First,
as $r$ is reduced
from $\infty$, $D(r)$
remains small as long as
$\Psi_{{\rm diag}}(r)$
and $\Psi_{{\rm exact}}(r)$ overlap. Then,
$\Psi_{{\rm diag}}(r)$ departs from
$\Psi_{{\rm exact}}(r)$ hence $D(r)$ reaches one or more maxima
at certain values of $r$ and eventually it vanishes again at the origin.
Therefore,
we define the real-space cutoff $r_c$ as the 
largest distance at which
\begin{equation}
D(r_c)=D_{{\rm threshold}}, 
\end{equation}
with  $D_{{\rm threshold}}$
being conventionally fixed to a few percentages.
The values of $r_c$ as a function of $N_b$ for the states illustrated
in figure \ref{f:cutoffs}(a) (attractive; $D_{{\rm threshold}}=0.02$)
and figure \ref{f:cutoffs}(b) (repulsive; $D_{{\rm threshold}}=0.01$),
respectively, are collected in table \ref{table2} (note that in the 
attractive case the convergence 
is more demanding and we therefore choose a more tolerant threshold value).
\begin{table}
\caption{\label{table2} Real-space cutoff, $r_c$,
vs number of oscillator orbitals in relative coordinates, $N_b$, 
employed in the direct 
diagonalization of the two-body ground-state
wave function $\Psi_{{\rm diag}}(r)$. 
Data of second and third columns correspond to energies 
$\mathcal{E}=1.500$ and $\mathcal{E}=2.500$,
being the largest roots of the
equation $D(r_c)=D_{{\rm threshold}}$, with $D_{{\rm threshold}}=0.02$
and 0.01, respectively.
The last column reports the first zero of the $N_b$th
noninteracting orbital along the radial direction as a reference.}
\begin{indented}
\item[]\begin{tabular}{cccc}
\br
$N_b$ & $\mathcal{E}=1.500$ ($g<0$) & $\mathcal{E}=2.500$ ($g>0$) 
& $\mathcal{E}=2$ ($g=0$) \\ 
\mr
2 & 2.58 & 4.54 & 2.41 \\
3 & 2.21 & 3.57 & 2.08 \\
4 & 2.00 & 3.14 & 1.91 \\
5 & 1.86 & 2.86 & 1.80 \\
6 & 1.77 & 2.67 & 1.73 \\
10& 1.53 & 2.21 & 1.55 \\
13& 1.41 & 1.99 & 1.48 \\
\br
\end{tabular}
\end{indented}
\end{table}
As expected, the real-space cutoff $r_c$ decreases
with the increasing energy cutoff $N_b$. 
As a reference, in table \ref{table2} we
also report the first zero of the oscillator orbitals with $n\le N_b$ 
in relative coordinates. All data show similar
trends and may be fitted by a power law of the type
$r_c \propto N_b^{\beta}$, where $\beta\approx -0.7 $ for
$\mathcal{E}=1.500$ as well as for $\mathcal{E}=2.500$, 
and $\beta \approx -0.6$ for
the noninteracting case $\mathcal{E}=2$. 
These exponents $\beta$ are not far from the value
$\beta = -0.5$, hence roughly
$N_b\sim 1/r_c^2$, i.e., the cutoffs 
in real and energy spaces are complementary
(assuming that the cutoff respectively
in momentum space, $p_c$, and in real space, $r_c$, are related
by $p_c\sim 1/r_c$).

\section{$N$ interacting particles in a trap}

So far we have considered the special case of two interacting 
particles for which the relative motion can be uncoupled from
the center-of-mass motion. In this case  
the results for two spin-1/2 fermions forming a singlet were the same as
for two spinless bosons. However, for $N>2$
this transformation turns out to be cumbersome for an
efficient implementation of the CI algorithm.

\begin{figure}[htbp]
\vspace{-1cm}
\setlength{\unitlength}{1 cm}
\begin{indented}
\item[]\begin{picture}(8.5,6.5)
\put(0.2,-0.2){\epsfig{file=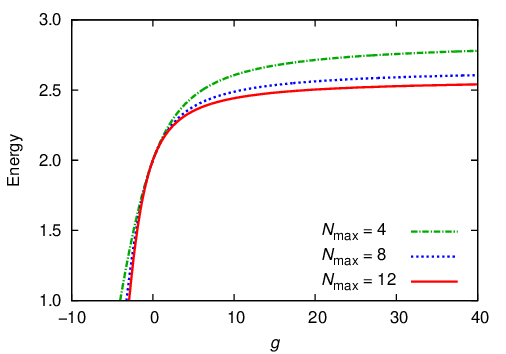,width=3.1in, ,angle=0}}
\end{picture}
\end{indented}
\caption{(color online)
Ground state energy $\mathcal{E}$  
of the two-particle system vs the coupling constant $g$. 
The different lines correspond to different numbers, $N_{{\rm max}}$, 
of harmonic oscillator shells employed in the CI diagonalization.
}
\label{f:Energies_02_gs_g}
\end{figure}

The Hamiltonian of $N$ particles in a harmonic trap 
is written in terms of the standard coordinates $\bm{r}_i$ as 
\begin{equation}
H_N = \sum_{i=1}^{N}\left(-\frac{1}{2}\nabla_i^2+\frac{1}{2}
r_i^2\right)+g\sum_{i<j}\delta(\bm{r}_i-\bm{r}_j). 
\label{eq:Hdimless}
\end{equation}
In typical CI codes the $N$-body Hilbert space is spanned
over a basis set of Slater determinants for fermions \cite{McWeeny,Helgaker,Jensen,Brasken2000,ReimannRMP,RontaniCI} or
permanents for bosons \cite{Sundholm2004,Szabados2012}.
The size of the Hilbert space for the $N$-body problem is 
determined by the number of 
single-particle orbitals from which the Fock states are built. 
In the 2D harmonic oscillator these single-particle orbitals 
are grouped in shells according to their single-particle energy.
One usually restricts either the noninteracting configurational energy, 
or simply only includes   
oscillator shells to a maximum number, $N_{{\rm max}}$.  
Here we choose the latter in order to keep the dimensionality 
well-defined.  
In the given space and for a given size of the interaction strength $g$ 
one then calculates the ground-state energy in the two-body system, 
$\mathcal{E}=E_{\rm GS}^{\rm CI}(g,N_{{\rm max}})$. 
The dependence of $\mathcal{E}$ on the interaction strength parameter 
is illustrated in figure \ref{f:Energies_02_gs_g} for a few different 
basis sizes. 

\begin{figure}[!ht]
\setlength{\unitlength}{1 cm}
\begin{indented}
\item[]\begin{picture}(8.5,7.0)
\put(0.2,-0.2){\epsfig{file=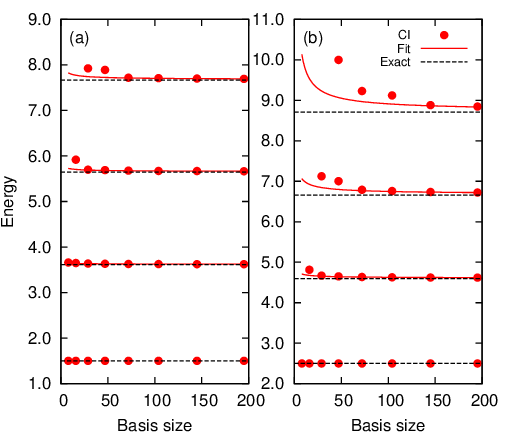,width=3.1in, ,angle=0}}
\end{picture}
\end{indented}
\caption{(color online)
Convergence of the four lowest energy states 
for two unpolarized fermions versus the many-body basis size 
in the CI diagonalization. The total angular momentum is
$M=0$ and calculations are performed for 
$N_{{\rm max}}=3$ to $10$. The interaction strength $g$ 
is determined to give an energy for the two-particle system of 
$\mathcal{E}=1.500$ (a) and $\mathcal{E}=2.500$ (b). 
Solid lines are power-law fits to the data and the 
dashed lines show the exact values given by (\ref{eq:nosbusch}). }
\label{f:Energies_02}
\end{figure}

For the $N=2$ system, performing the direct CI calculation  
is alternative to the relative-coordinate approach considered 
in Section \ref{s:two}. The energy levels up to the third excited 
state are calculated, and figure \ref{f:Energies_02} shows the convergence 
for each of the states as the basis size is increased. 
Here we observe a fast convergence towards the exact values 
which is well described by a power-law decay in the basis size. 
The higher-energy states have wave function components involving 
higher shells and need a larger basis size to converge. 

\subsection{Energy spectrum of three fermions}
\label{s:threeA}

\begin{figure}
\vspace{9mm}
\begin{indented}
\item[] \centerline{\epsfig{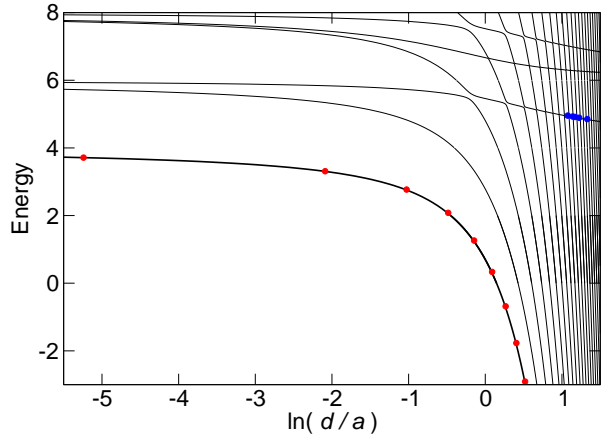}}
\end{indented}
\caption{(color online)  
Exact (solid lines)
and CI (bullets) energy spectrum of three unpolarized
fermions vs scattering length $a$ for angular momentum
$M=1$. Red [light gray] and blue [dark gray] bullets correspond 
to attractive ($g<0$) and repulsive ($g>0$) interactions, respectively.
Here $d=\sqrt{2}$ is the harmonic oscillator length in the 
relative-motion frame.
In the CI calculations we used $N_{\rm max}=10$.
\label{three_fermions}}
\end{figure}

A  semi-analytic solution is available also for three 
interacting particles in a harmonic trap \cite{Drummond}.
Figure \ref{three_fermions} compares the exact energies 
of $N=3$ unpolarized fermions 
(solid lines) with those obtained by the CI method (red [light gray]
and blue [dark gray] bullets,
respectively, for $g<0$ and $g>0$). The exact energies are obtained following
the method reported in reference \cite{Drummond} (there are a few discrepancies with respect to our
derivation \footnote{Equation (7) in reference \cite{Drummond} 
should be replaced with our equation 
(\ref{eq:nosbusch}).  Moreover, equation (17) should 
read $B_n=(-1)^m\left[\psi(-\nu_{m,n}) -2 \,{\rm ln}(d/a)\right]$.}).
Here we focus on the part of the spectrum with 
angular momentum $M=1$, since it also provides the ground state, and plot the
energies as a function of ${\rm ln}(d/a)$, with $d=\sqrt{2}$ being the 
oscillator length in the relative frame. 
The ground-state energy $\mathcal{E}$ in the two-particle system 
is adjusted via the choice of the coupling strength, as described above, 
and (\ref{eq:nosbusch}) relates it to the scattering length. 
As the scattering length $a$ decreases from its infinite positive value
at $\ln(d/a) \rightarrow - \infty$, the ground-state energy of the $N=3$ system
departs from the noninteracting value  
$E=4$ and rapidly falls off. First, the CI energies of both
ground and excited states match the exact ones very well
(up to the fourth digit for the ground state at $\ln(d/a)=-5.240$
and third digit at $\ln(d/a)=-1.027$). As $\ln(d/a)$ increases, however,  
they progressively loose accuracy,  as illustrated in table \ref{t:3F}.  
\begin{table}
\caption{\label{t:3F} Comparison between exact and
CI ground-state energies of three unpolarized fermions 
for attractive 
interaction, as shown in figure \ref{three_fermions}
(thick solid line and red [light gray] bullets, respectively). 
In the CI calculations we used $N_{\rm max}=10$.
The values of the CI coupling constant $g$ and
corresponding scattering length $a$ are given.
}
\begin{indented}
\item[]\begin{tabular}{c|c|c|c}
\br
$\ln(d/a)$  & {\rm exact} & $g$& {\rm CI} \\ 
\mr
$-\infty$  & 4   &0     & 4.000  \\
 -5.240   & 3.712&-1   & 3.713  \\
 -2.089   & 3.305  &-2 & 3.311  \\
 -1.027   & 2.756 &-3  & 2.769  \\
 -0.4834  & 2.061 &-4  & 2.082  \\
 -0.1458  & 1.233 &-5  & 1.263  \\
  0.08946 & 0.2973&-6  & 0.3338 \\
  0.2660  & -0.7261&-7 & -0.6832 \\
  0.4055  & -1.817 &-8 & -1.769  \\
  0.5197  & -2.963 &-9 & -2.910  \\
\br
\end{tabular}
\end{indented}
\end{table}
The increasing discrepancy is a consequence
of the fact that the
energy cutoff induces a real-space cutoff that prevents 
resolving the wave function when the interparticle separation is too 
small. This is in particular  
the case for those low-lying states whose wave functions 
spatially collapse into strongly bound molecular trimers.  

The region ${\rm ln}(d/a) > 0.5$ of figure \ref{three_fermions} 
is considerably more complex, since now two different types
of excitations appear: (i) a dense set of ground-state replicas that are
tighly bound molecular dimers plus a third spectator
particle in an excited level, whose energies 
have almost vertical slopes, and  
(ii) a discrete set of curves with moderate slopes that may be regarded
as the levels of three fermions with repulsive interactions \cite{Drummond}.
Whereas there is no clear-cut distinction between the two types of
curves due to the many avoided crossings, the CI method is obviously able
to reproduce only levels (ii) with significant accuracy. 
States (i) are clearly 
beyond the reach of the technique, as the radius of tightly bound dimers
is smaller than the mimimum spatial resolution associated with the Hilbert 
space size provided by $N_{\rm max}$.

\subsection{Energy spectrum of three bosons}
\label{s:threeB}

\begin{figure}
\vspace{9mm}
\begin{indented}
\item[] \centerline{\epsfig{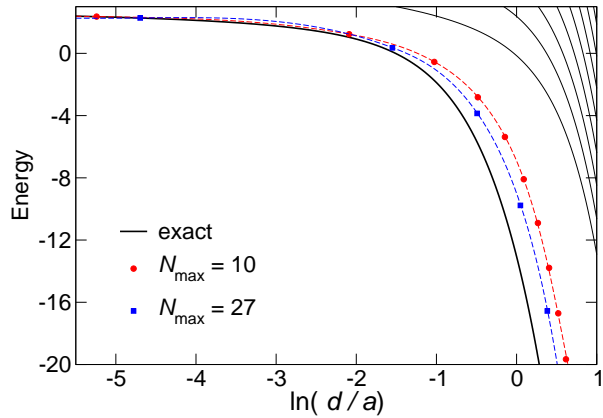}}
\end{indented}
\caption{(color online)  
Exact (solid lines)
and CI (bullets) energies of three spinless bosons
vs scattering length $a$ for total angular momentum $M=0$.
Here $d=\sqrt{2}$ is the harmonic oscillator length
in the relative-motion frame.
Respectively, $N_{{\rm max}}=10$ (circles) and 
$N_{{\rm max}}=27$ (squares) oscillator
shells were employed in the CI calculation.
The dashed lines are cubic smoothing spline interpolations
to CI results.
\label{three_bosons}}
\end{figure}

A plot similar to figure \ref{three_fermions}
is shown in figure \ref{three_bosons} for three spinless bosons,
except that the GS total angular momentum is now $M=0$.
In this case the accuracy of the CI calculation is significantly lower
than that in figure \ref{three_fermions}, at least
for the lowest-lying branch
of strongly bound trimers. The computational bottleneck is 
the absence of the short-range Pauli repulsion
that originates from the exchange between fermions.  
Even if an increase of the maximum number of oscillator shells 
$N_{{\rm max}}$
used for the single-particle basis set in the CI diagonalization
significantly improves
the matching between exact (thick solid line) and CI data (bullets),
going from $N_{{\rm max}}=10$ (circles) to
$N_{{\rm max}}=27$ (squares), the absolute 
error on the energy is as large as $\approx 2$ at $\ln(d/a)\approx -0.5$
and $\approx 5$ at $\ln(d/a)\approx 0$.

\subsection{Larger systems}\label{s:larger}

For the few-body systems with two and three particles treated so far, 
exact solutions exist and could be compared to.  
For larger systems, however, 
one must solely rely on numerical calculations. 
\begin{figure}[htbp]
\setlength{\unitlength}{1 cm}
\begin{indented}
\item[] \begin{picture}(8.5,7.0)
\put(0.2,-0.2){\epsfig{file=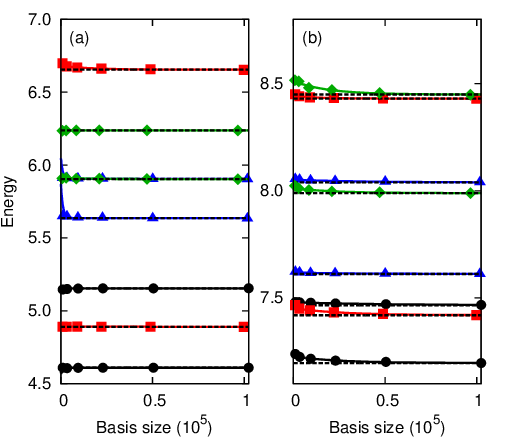,width=3.1in, ,angle=0}}
\end{picture}
\end{indented}
\caption{(color online)
Convergence of the energy spectrum for $N=4$ unpolarized 
fermions vs many-body basis size in the CI diagonalization. 
The coupling constant $g$ is determined to give an energy for 
the two-particle system of $\mathcal{E}=1.500$ (a) and 
$\mathcal{E}=2.500$ (b). 
Here, the two lowest energy states for each of the four total angular 
momenta $M=0,1,2$ or $3$ (circles, triangles, squares, and diamonds, 
respectively) are shown. Data points are computed for $N_{{\rm max}}=5$
to 10. Solid lines are exponential fits to the data and 
dashed lines are their asymptotic values.
}
\label{f:Energies_04}
\end{figure}

\begin{figure}[htbp]
\setlength{\unitlength}{1 cm}
\begin{indented}
\item[]\begin{picture}(8.5,7.0)
\put(0.2,-0.2){\epsfig{file=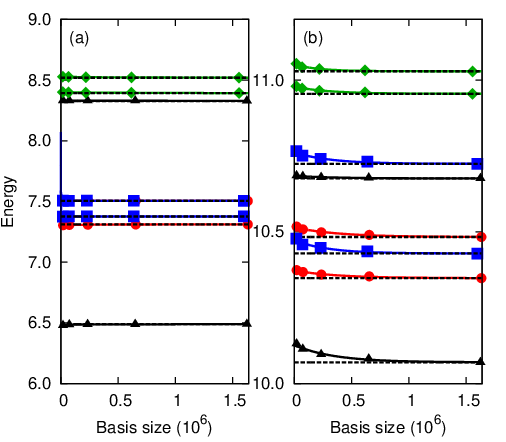,width=3.1in, ,angle=0}}
\end{picture}
\end{indented}
\caption{(color online)
Same as figure \ref{f:Energies_04} but for $N=5$ unpolarized fermions. }
\label{f:Energies_05}
\end{figure}
The many-body Hamiltonian given in (\ref{eq:Hdimless}) 
is diagonalized in a basis of Slater determinants constructed 
from the lowest $N_{\rm max}$ harmonic oscillator shells. 
In figures \ref{f:Energies_04} and \ref{f:Energies_05}, 
for the example of a system of respectively four and five 
unpolarized fermions, we show how the energies depend on 
the size of the Hilbert space used in the CI calculation. 
Utilizing the discussed concept of renormalization the 
coupling constant $g$ together with the basis size are 
chosen to give a GS energy of the two-particle system of 
either $\mathcal{E}=1.500$ (attractive interactions) or 
$\mathcal{E}=2.500$ (repulsive interactions). We show the two lowest 
energy states for the four angular momenta $M=0,1,2$ and $3$. 
For the parameters chosen we find that the energies of low-lying states 
are well converged for manageable sizes of the Hilbert space.

\section{Conclusions}
\label{s:four}

The aim of this paper was to provide a practical ground for CI 
calculations of $N$ particles interacting via a contact potential 
that needs to be properly regularized. To achieve this in a simple 
and straightforward manner, one  renormalizes the strength of the 
contact potential for two particles in a given subspace of 
single-particle basis states. The final outcome is that the CI 
diagonalization over a finite basis set provides physically 
relevant observables; the energy cutoff only affects the resolution 
on complementary real space distances, while the low-lying
excitation spectrum is unaffected.
The  procedure discussed here relies on the comparison to both the 
energy and wave function of two and three particles, obtained in two 
different ways. The first way is the CI diagonalization with an energy 
cutoff, while the second one is the exact solution of the 
Schr\"odinger equation for the regularized form of the contact 
pseudopotential. Both ground and excited states are considered in 
this comparison. The analysis of CI data for a truncated Hilbert space 
provides a fully consistent physical picture of the results as well as 
a systematic assessment of the error of the calculation. Finally, 
the scalability of the method was demonstrated for larger fermionic 
systems with $N=4$ and $N=5$ as an example, where no analytical 
solutions exist and one must rely on numerical calculations. 
The method converges well for fermionic few-body systems with attractive 
as well as repulsive interaction, while the bosonic case is found 
to be more cumbersome. We expect that this procedure may be applied 
to $N>5$, provided that the radius of studied few-body complexes
is larger than the spatial resolution associated
with the size of the truncated Hilbert space.

The method discussed here validates previous work, where such 
\emph{ad hoc} renormalization by simple Hilbert space truncation 
has been applied, 
see for example, the discussion of pairing and shell structure in 
finite-size fermion systems \cite{Rontani09}, or the recent 
analysis of  few- to many-body transition and the Higgs mode 
in a paired Fermi gas \cite{Bjerlin2015}.
The scheme discussed here may be useful for future diagonalization 
studies in finite-size fermion systems that have now also become 
experimentally accessible \cite{Serwane2011,Zurn2013,Wenz2013}.

\ack
We thank Jonas Cremon for discussions and for his assistance with 
part of the numerical work at an early stage of the project, 
as well as for providing the boson data in Figure 7.  
We also thank Georg Bruun, Daniela Pfannkuche, Chris Pethick, 
Ben Mottelson, Vladimir Zelevinsky, and Pavel Kurasov for discussions. 
This work was supported by EU-FP7 Marie Curie ITN INDEX,
by Fondazione Cassa di Risparmio di Modena
through the project COLDandFEW, by the CINECA-ISCRA grant 
IscrC\_TUN1DFEW,
as well as by NordForsk, the Swedish Foundation
for Strategic Research, the Swedish Research Council, and
NanoLund at Lund University.   

\section*{References}
\bibliography{deltareg}

\begin{thebibliography}{10}

\bibitem{Dalfovo1999}
F.~Dalfovo, S.~Giorgini, L.~P. Pitaevskii, and S.~Stringari.
\newblock Theory of {B}ose-{E}instein condensation in trapped gases.
\newblock {\em Rev. Mod. Phys.}, 71:463--512, 1999.

\bibitem{Leggett2001}
A.~J. Leggett.
\newblock {B}ose-{E}instein condensation in the alkali gases: {S}ome
  fundamental concepts.
\newblock {\em Rev. Mod. Phys.}, 73:307--356, 2001.

\bibitem{PethickSmith2002}
C.~J. Pethick and H.~Smith.
\newblock {\em Bose-Einstein Condensation in Dilute Gases}.
\newblock Cambridge University Press, Cambridge (UK), 2002.

\bibitem{PitaevskiiStringari2003}
L.~Pitaevskii and S.~Stringari.
\newblock {\em Bose-{E}instein {C}ondensation}.
\newblock Oxford University Press, Oxford, 2003.

\bibitem{Leggett2006}
A.~J. Leggett.
\newblock {\em Quantum Liquids}.
\newblock Oxford University Press, Oxford, 1st edition, 2006.

\bibitem{Bloch2008}
I.~Bloch, J.~Dalibard, and W.~Zwerger.
\newblock Many-body physics with ultracold gases.
\newblock {\em Rev. Mod. Phys.}, 80:885--964, 2008.

\bibitem{Giorgini2008}
S.~Giorgini, L.~P. Pitaevskii, and S.~Stringari.
\newblock Theory of ultracold atomic {F}ermi gases.
\newblock {\em Rev. Mod. Phys.}, 80:1215--1274, 2008.

\bibitem{Huang}
K.~Huang.
\newblock {\em Statistical Mechanics}.
\newblock Wiley, New York, 1963.

\bibitem{Galitskii1958}
V.~M. Galitskii.
\newblock The energy spectrum of a non-ideal {F}ermi gas.
\newblock {\em Soviet Phys. JETP}, 34:104--112, 1958.

\bibitem{Gorkov1961}
L.~P. Go{r'k}ov and T.~K. Melik-Barkhudarov.
\newblock Contribution to the theory of superfluidity in an imperfect {F}ermi
  gas.
\newblock {\em Soviet Phys. JETP}, 13:1018--1022, 1961.

\bibitem{Leggett1980bis}
A.~J. Leggett.
\newblock Diatomic molecules and {C}ooper pairs.
\newblock In A.~Pekalski and A.~Przystawa, editors, {\em Modern Trends in the
  Theory of Condensed Matter}, volume 115 of {\em Lecture Notes in Physics},
  chapter~2, pages 13--27. Springer-Verlag, Berlin, 1980.

\bibitem{adhikari1997}
S.~Adhikari and A.~Ghosh.
\newblock Renormalization in non-relativistic quantum mechanics.
\newblock {\em J. Phys. A: Math. Gen.}, 30:6553, 1997.

\bibitem{ghosh1998}
A.~Ghosh, S.~Adhikari, and B.~Talukdar.
\newblock Dimensional versus cut-off renormalization and the nucleon-nucleon
  interaction.
\newblock {\em Phys. Rev. C}, 58:1913--1920, 1998.

\bibitem{Camblong2002}
H.~E. Camblong and C.~R. Ord{\'o\~n}ez.
\newblock Renormalized path integral for the two-dimensional $\delta$-function
  interaction.
\newblock {\em Phys. Rev. A}, 65:052123, 2002.

\bibitem{Pricoupenko06}
L.~Pricoupenko.
\newblock Pseudopotential in resonant regimes.
\newblock {\em Phys. Rev. A}, 73:012701, 2006.

\bibitem{suto}
A.~S{\"u}t{\'o}.
\newblock Exact eigenstates for contact interactions.
\newblock {\em J. Stat. Phys.}, 109:1051--1072, 2002.

\bibitem{Busch}
T.~Busch, {B.-G}. Englert, K.~Rz{\c{a}\.z}ewski, and M.~Wilkens.
\newblock Two cold atoms in a trap.
\newblock {\em Found. Phys.}, 28:549--559, 1998.

\bibitem{Cabo}
A.~Cabo, J.~L. Lucio, and H.~Mercado.
\newblock On scale invariance and anomalies in quantum mechanics.
\newblock {\em Am. J. Phys.}, 66:240--246, 1998.

\bibitem{Esry}
B.~D. Esry and C.~H. Greene.
\newblock Validity of the shape-independent approximation for {B}ose-{E}instein
  condensates.
\newblock {\em Phys. Rev. A}, 60:1451--1462, 1999.

\bibitem{Castin}
Y.~Castin.
\newblock Bose-{E}instein condensates in atomic gases: {S}imple theoretical
  results.
\newblock In R.~Kaiser, C.~Westbrook, and F.~David, editors, {\em Coherent
  Atomic Matter Waves}, Lectures Notes of Les Houches Summer School, pages
  1--136. EDP Sciences and Springer-Verlag, Berlin, 2001.

\bibitem{Pricoupenko2011}
L.~Pricoupenko.
\newblock Isotropic contact forces in arbitrary representation: Heterogeneous
  few-body problems and low dimensions.
\newblock {\em Phys. Rev. A}, 83:062711, 2011.

\bibitem{Doganov2013}
R.~A. Doganov, S.~Klaiman, O.~E. Alon, A.~I. Streltsov, and L.~S. Cederbaum.
\newblock Two trapped particles interacting by a finite-range two-body
  potential in two spatial dimensions.
\newblock {\em Phys. Rev. A}, 87:033631, 2013.

\bibitem{Blume2007}
D.~Blume, J.~von Stecher, and C.~H. Greene.
\newblock Universal properties of a trapped two-component {F}ermi gas at
  unitarity.
\newblock {\em Phys. Rev. Lett.}, 99:233201, 2007.

\bibitem{Stecher2007bis}
J.~von Stecher and C.~H. Greene.
\newblock Spectrum and dynamics of the {BCS-BEC} crossover from a few-body
  perspective.
\newblock {\em Phys. Rev. Lett.}, 99:090402, 2007.

\bibitem{Stecher2007ter}
J.~von Stecher, C.~H. Greene, and D.~Blume.
\newblock {BEC-BCS} crossover of a trapped two-component {F}ermi gas with
  unequal masses.
\newblock {\em Phys. Rev. A}, 76:053613, 2007.

\bibitem{vonStecher2008}
J.~von Stecher, C.~H. Greene, and D.~Blume.
\newblock Energetics and structural properties of trapped two-component {F}ermi
  gases.
\newblock {\em Phys. Rev. A}, 77:043619, 2008.

\bibitem{Christensson2009}
J.~Christensson, C.~Forss{\'e}n, S.~{\AA}berg, and S.~M. Reimann.
\newblock Effective-interaction approach to the many-boson problem.
\newblock {\em Phys. Rev. A}, 79:012707, 2009.

\bibitem{McWeeny}
R.~McWeeny.
\newblock {\em Methods of {M}olecular {Q}uantum {M}echanics}.
\newblock Academic Press, London, 1992.

\bibitem{Helgaker}
T.~Helgaker, P.~J{\o}rgensen, and J.~Olsen.
\newblock {\em Molecular {E}lectronic-{S}tructure {T}heory}.
\newblock Wiley, Chichester, England, 2000.

\bibitem{Jensen}
F.~Jensen.
\newblock {\em Introduction to {C}omputational {C}hemistry}.
\newblock Wiley, Chichester, England, 2007.

\bibitem{Brasken2000}
M.~Brasken, M.~Lindberg, D.~Sundholm, and J.~Olsen.
\newblock Full configuration interaction calculations of electron-hole
  correlation effects in strain-induced quantum dots.
\newblock {\em Phys. Rev. B}, 61:7652--7655, 2000.

\bibitem{ReimannRMP}
S.~M. Reimann and M.~Manninen.
\newblock Electronic structure of quantum dots.
\newblock {\em Rev. Mod. Phys.}, 74:1283--1342, 2002.

\bibitem{RontaniCI}
M.~Rontani, C.~Cavazzoni, D.~Bellucci, and G.~Goldoni.
\newblock Full configuration interaction approach to the few-electron problem
  in artificial atoms.
\newblock {\em J. Chem. Phys.}, 124:124102, 2006.

\bibitem{Fetter2009}
A.~Fetter.
\newblock Rotating trapped {B}ose-{E}instein condensates.
\newblock {\em Rev. Mod. Phys.}, 81:647--691, 2009.

\bibitem{Castinbis}
Y.~Castin.
\newblock Simple theoretical tools for low dimension {B}ose gases.
\newblock {\em J. Phys. IV}, 116:89--132, 2004.

\bibitem{Drut2013}
J.~E. Drut and A.~N. Nicholson.
\newblock Lattice methods for strongly interacting many-body systems.
\newblock {\em J. Phys. G: Nucl. Part. Phys.}, 40:043101, 2013.

\bibitem{Braaten1997}
E.~Braaten and A.~Nieto.
\newblock Quantum corrections to the ground state of a trapped
  {B}ose-{E}instein condensate.
\newblock {\em Phys. Rev. B}, 56:14745--14765, 1997.

\bibitem{Bulgac2006}
A.~Bulgac, J.~E. Drut, and P.~Magierski.
\newblock Spin 1/2 fermions in the unitary regime: {A} superfluid of a new
  type.
\newblock {\em Phys. Rev. Lett.}, 96:090404, 2006.

\bibitem{Stecher2007}
J.~von Stecher and C.~H. Greene.
\newblock Renormalized mean-field theory for a two-component {F}ermi gas with
  $s$-wave interactions.
\newblock {\em Phys. Rev. A}, 75:022716, 2007.

\bibitem{Stetcu2007a}
I.~Stetcu, B.~R. Barrett, and U.~van Kolck.
\newblock No-core shell model in an effective-field-theory framework.
\newblock {\em Phys. Lett. B}, 653:358--362, 2007.

\bibitem{Stetcu2007b}
I.~Stetcu, B.~R. Barrett, U.~van Kolck, and J.~P. Vary.
\newblock Effective theory for trapped few-fermion systems.
\newblock {\em Phys. Rev. A}, 76:063613, 2007.

\bibitem{Alhassid2008}
Y.~Alhassid, G.~F. Bertsch, and L.~Fang.
\newblock New effective interaction for the trapped {F}ermi gas.
\newblock {\em Phys. Rev. Lett.}, 100:230401, 2008.

\bibitem{Zinner2009}
N.~T. Zinner, K.~M{\o}lmer, C.~{\"O}zen, D.~J. Dean, and K.~Langanke.
\newblock Shell-model {M}onte {C}arlo simulations of the {BCS-BEC} crossover in
  few-fermion systems.
\newblock {\em Phys. Rev. A}, 80:013613, 2009.

\bibitem{Gilbreth2012}
C.~N. Gilbreth and Y.~Alhassid.
\newblock Separable effective interaction for the trapped {F}ermi gas: {T}he
  {BEC-BCS} crossover.
\newblock {\em Phys. Rev. A}, 85:033621, 2012.

\bibitem{Tolle2013}
S.~T{\"o}lle, {H.-W}. Hammer, and B.~C. Metsch.
\newblock Convergence properties of the effective theory for trapped bosons.
\newblock {\em J. Phys. G: Nucl. Part. Phys.}, 40:055004, 2013.

\bibitem{Gilbreth2013}
C.~N. Gilbreth and Y.~Alhassid.
\newblock Pair condensation in a finite trapped {F}ermi gas.
\newblock {\em Phys. Rev. A}, 88:063643, 2013.

\bibitem{Bulgac2002}
A.~Bulgac and Y.~Yu.
\newblock Renormalization of the {H}artree-{F}ock-{B}ogoliubov equations in the
  case of a zero range pairing interaction.
\newblock {\em Phys. Rev. Lett.}, 88:042504, 2002.

\bibitem{Pricoupenko2004}
L.~Pricoupenko.
\newblock Variational approach for the two-dimensional trapped
  {B}ose-{E}instein condensate.
\newblock {\em Phys. Rev. A}, 70:013601, 2004.

\bibitem{Johnson2010}
C.~W. Johnson.
\newblock Many-body fits of phase-equivalent effective interactions.
\newblock {\em Phys. Rev. C}, 82:031303(R), 2010.

\bibitem{Boettcher2012}
I.~Boettcher, J.~M. Pawlowski, and S.~Diehl.
\newblock Ultracold atoms and the functional renormalization group.
\newblock {\em Nucl. Phys. B (Proc. Suppl.)}, 228:63--135, 2012.

\bibitem{Yan2015}
Y.~Yan and D.~Blume.
\newblock Incorporating exact two-body propagators for zero-range interactions
  into {$N$}-body {M}onte {C}arlo simulations.
\newblock {\em Phys. Rev. A}, 91:043607, 2015.

\bibitem{Drummond}
{X.-J}. Liu, H.~Hu, and P.~D. Drummond.
\newblock Three attractively interacting fermions in a harmonic trap: {E}xact
  solution, ferromagnetism, and high-temperature thermodynamics.
\newblock {\em Phys. Rev. B}, 82:054524, 2010.

\bibitem{Olshanii}
M.~Olshanii and L.~Pricoupenko.
\newblock Rigorous approach to the problem of ultraviolet divergencies in
  dilute bose gases.
\newblock {\em Phys. Rev. Lett.}, 88:010402, 2002.

\bibitem{Petrov2001}
D.~S. Petrov and G.~V. Shlyapnikov.
\newblock Interatomic collisions in a tightly confined {B}ose gas.
\newblock {\em Phys. Rev. A}, 64:012706, 2001.

\bibitem{Coon2012}
S.~A. Coon, M.~I. Avetian, M.~K.~G. Kruse, U.~van Kolck, P.~Maris, and J.~P.
  Vary.
\newblock Convergence properties of ab initio calculations of light nuclei in a
  harmonic oscillator basis.
\newblock {\em Phys. Rev. C}, 86:054002, 2012.

\bibitem{Konig2014}
S.~K{\"o}nig, S.~K. Bogner, R.~J. Furnstahl, S.~N. More, and T.~Papenbrock.
\newblock Ultraviolet extrapolations in finite oscillator bases.
\newblock {\em Phys. Rev. C}, 90:064007, 2014.

\bibitem{Sundholm2004}
D.~Sundholm and T.~V{\"{a}}nsk{\"{a}}.
\newblock A configuration interaction approach to bosonic systems.
\newblock {\em J. Phys. B: At. Mol. Opt. Phys.}, 37:2933--2942, 2004.

\bibitem{Szabados2012}
A.~Szabados, P.~Jeszenszki, and P.~R. Surj{\'a}n.
\newblock Efficient iterative diagonalization of the {B}ose-{H}ubbard model for
  ultracold bosons in a periodic optical trap.
\newblock {\em Chem. Phys.}, 401:208--216, 2012.

\bibitem{Rontani09}
M.~Rontani, J.~R. Armstrong, Y.~Yu, S.~{\AA}berg, and S.~M. Reimann.
\newblock Cold fermionic atoms in two-dimensional traps: {P}airing versus
  {H}und’s rule.
\newblock {\em Phys. Rev. Lett.}, 102:060401, 2009.

\bibitem{Bjerlin2015}
J.~Bjerlin, S.~M. Reimann, and G.~M. Bruun.
\newblock Few-boy precursor of the {H}iggs mode in a {F}ermi gas.
\newblock {\em Phys. Rev. Lett.}, 116:155302, 2016.

\bibitem{Serwane2011}
F.~Serwane, G.~Z{\"u}rn, T.~Lompe, T.~B. Ottenstein, A.~N. Wenz, and S.~Jochim.
\newblock Deterministic preparation of a tunable few-fermion system.
\newblock {\em Science}, 332:336--338, 2011.

\bibitem{Zurn2013}
G.~Z{\"u}rn, A.~N. Wenz, S.~Murmann, A.~Bergschneider, T.~Lompe, and S.~Jochim.
\newblock Pairing in few-fermion systems with attractive interactions.
\newblock {\em Phys. Rev. Lett.}, 111:175302, 2013.

\bibitem{Wenz2013}
A.~N. Wenz, G.~Z{\"u}rn, S.~Murmann, I.~Brouzos, T.~Lompe, and S.~Jochim.
\newblock From few to many: {O}bserving the formation of a {F}ermi sea one atom
  at a time.
\newblock {\em Science}, 342:457--460, 2013.

\end{thebibliography}
\bibliographystyle{unsrt}
\end{document}